\documentclass[conference]{IEEEtran}
\usepackage{times}

\usepackage{physics}
\usepackage{hyperref}
\usepackage{graphicx}
\usepackage[font=footnotesize]{caption}
\usepackage{amsmath}
\usepackage{amsfonts}
\usepackage{amssymb}
\usepackage{amsthm}
\usepackage{tabularx}
\usepackage{mathrsfs} 
\usepackage{cite}

\usepackage{subfig}

\usepackage{soul}

\usepackage{stfloats}
\usepackage{float}
\usepackage{url}
\usepackage{hyperref}

\usepackage[nolist,nohyperlinks]{acronym}
\usepackage[numbers,sort&compress]{natbib}

\usepackage{bbm}

\usepackage{array}
\usepackage{booktabs}
\usepackage{colortbl}
\usepackage{mdwmath}
\usepackage{mdwtab}
\usepackage{eqparbox}
\usepackage{cellspace}
\usepackage{makecell}

\DeclareGraphicsExtensions{.png,.eps,.ps,.pdf}

\usepackage{tikz}
\usetikzlibrary{arrows,    
                shapes, 
                positioning, 
                fit,    
                backgrounds, 
                mindmap,
                petri,
                intersections, 
                external %
                }
\definecolor{emerald}{RGB}{69,155,61}
\definecolor{gold}{RGB}{244,216,51}
\definecolor{pink}{RGB}{235,44,206}
\usepackage{verbatim}

\begin{acronym}
\acro{AI}{Artificial Intelligence}
\acro{AMC}{adaptive modulation and coding}
\acro{AWGN}{additive white Gaussian noise}
\acro{BLER}{block error rate}
\acro{CNN}{convolutional neural network}
\acro{RNN}{recurrent neural network}
\acro{SNR}{signal-to-noise ratio}
\acrodefplural{SNR}[SNRs]{signal-to-noise ratios}
\acro{RV}{random variable}
\acrodefplural{RV}[RVs]{random variables}
\acro{NLOS}{non line-of-sight}
\acro{LOS}{line-of-sight}
\acrodefplural{LoS}{line-of-sight}
\acro{EH}{Energy Harvesting}
\acro{PDF}{probability density function}
\acro{CDF}{cumulative distribution function}
\acrodefplural{CDF}[CDFs]{cumulative distribution functions}
\acro{IoT}{Internet of Things}
\acro{PB}{Power Beacon}
\acrodefplural{PB}[PBs]{Power Beacons}
\acro{WPC}{Wireless Powered Communications}
\acro{WPT}{Wireless Power Transmission}
\acro{RFID}{Radio Frequency IDentification}
\acro{MC}{Monte Carlo}
\acro{PLS}{Physical Layer Security}
\acro{SWIPT}{Simultaneous Wireless and Information Power Transfer}
\acro{MIMO}{Multiple-Input Multiple-Output}
\acro{OFDM}{Orthogonal Frequency Division Multiplexing}
\acro{CQI}{channel quality indicator}
\acro{PMI}{precoding matrix indicator}
\acro{RI}{rank indicator}
\acro{LA}{link adaptation}
\acro{CSI}{channel state information}
\acro{TDD}{time division duplexing}
\acro{FDD}{frequency division duplexing}
\acro{FLOPs}{floating point operations per second}
\acro{RB}{resource block}
\acrodefplural{RB}[RBs]{resource blocks}
\acro{RE}{resource element}
\acrodefplural{RE}[REs]{resource elements}
\acro{BS}{base station}
\acrodefplural{BS}[BSs]{base stations}
\acro{UE}{user equipment}
\acrodefplural{UE}[UEs]{user equipments}
\acro{LEO}{Low Earth Orbit}
\acro{TDL}{tapped delay line}
\acro{NMSE}{normalized mean squared error}
\acro{MCS}{modulation and coding scheme}
\acro{SRS}{Sounding Reference Signal}
\acro{DNN}{deep neural network}
\acrodefplural{DNN}[DNNs]{deep neural networks}
\acro{RNN}{recurrent neural network}
\acrodefplural{RNN}[RNNs]{recurrent neural networks}
\acro{LSTM}{long short-term memory}
\acro{LS}{Least Square}
\acro{SINR}{signal to interference plus noise ratio }
\acro{EESM}{exponential effective SINR mapping}
\end{acronym}

\tikzstyle{int}=[draw, fill=cyan!20, minimum size=2em]
\tikzstyle{int_blue}=[draw, fill=black!20, minimum size=2em]
\tikzstyle{int_green}=[draw, fill=green!20, minimum size=2em]
\tikzstyle{int_red}=[draw, fill=red!20, minimum size=2em]
\tikzstyle{init} = [pin edge={to-,thin,black}]

\begin{document}
\title{Deep Learning-Based CSI Prediction Framework for Channel Aging Mitigation in TDD 5G Systems}
\author{\IEEEauthorblockN{Francisco D\'iaz-Ruiz, Francisco J. Mart\'in-Vega, Jos\'e Antonio Cort\'es, Gerardo G\'omez \\ and Mari Carmen Aguayo-Torres}
\IEEEauthorblockA{Communications and Signal Processing Lab, Telecommunication Research Institute (TELMA)\\
Universidad de M\'alaga, Bulevar Louis Pasteur 35, 29010 M\'alaga (Spain) }
}
\thanks{Manuscript received April xx, 2025; revised XXX. %
%
This work has been funded by MCIN/AEI/10.13039/501100011033 (Spain), by the European Fund for Regional Development (FEDER) {`\textit{A way of making Europe}´}, Keysight Technologies, and the University of M\'alaga through grants PID2020-118139RB-I00, RYC2021-034620-I and 8.06/6.10.6635.}
\thanks{The authors are with the Communications and Signal Processing Lab, Telecommunication Research Institute (TELMA), Universidad de M\'alaga, E.T.S. Ingenier\'ia de Telecomunicaci\'on, Bulevar Louis Pasteur 35, 29010 M\'alaga (Spain). (e-mail: fdiaz@ic.uma.es)}


\maketitle
\begin{abstract}
\Ac{TDD} has become the dominant duplexing mode in 5G and beyond due to its ability to exploit channel reciprocity for efficient downlink \ac{CSI} acquisition. However, channel aging caused by user mobility and processing delays degrades the accuracy of CSI, leading to suboptimal link adaptation and loss of performance. To address this issue, we propose a learning-based CSI prediction framework that leverages temporal correlations in wireless channels to forecast future \ac{SINR} values. The prediction operates in the effective SINR domain, obtained via \ac{EESM}, ensuring full compatibility with existing 5G standards without requiring continuous pilot signaling.

Two models are considered: a fully connected \ac{DNN} and an \ac{LSTM}-based network. The simulation results show that the \ac{LSTM} predictor achieves an improvement of up to 2~dB in \ac{NMSE} and a gain of up to 1.2~Mbps throughput over a baseline without prediction under moderate Doppler conditions. These results confirm the potential of lightweight AI-based CSI prediction to effectively mitigate channel aging and enhance link adaptation in \ac{TDD} 5G systems.
\end{abstract}


\section{Introduction}

\IEEEPARstart{T}{ime-division duplexing} (TDD) has become the dominant duplexing mode in 5G and beyond, primarily due to its ability to exploit channel reciprocity for efficient downlink \acf{CSI} acquisition. By leveraging uplink pilots, \ac{TDD} reduces signalling overhead compared to \ac{FDD} and enables more agile \ac{LA} strategies. Through dynamic adjustment of transmission parameters such as the \ac{MCS}, \ac{LA} seeks to maximize spectral efficiency while ensuring reliable communication~\cite{Vega2021,Goldsmith2005}.

A major challenge in \ac{TDD} systems is \emph{channel aging}, which arises from the delay between channel estimation and its use in data transmission~\cite{Papazafeiropoulos2017}. User mobility and processing latency exacerbate this effect, leading to outdated \ac{CSI}, suboptimal \ac{MCS} selection, higher error rates, and throughput degradation~\cite{Truong2013}. While transmitting pilots more frequently can mitigate channel aging, this comes at the cost of increased signalling overhead, ultimately reducing system efficiency.

Predictive \ac{CSI} has emerged as a promising solution to this trade-off by leveraging the temporal correlation of wireless channels to forecast future states~\cite{Jiang2017}. Recent advances in artificial intelligence (AI) and machine learning (ML) have accelerated this trend: deep learning models such as convolutional and recurrent neural networks can capture complex, non-linear channel dynamics and have demonstrated strong performance in mobility-aware prediction and adaptive resource allocation~\cite{Ye2018,Liao2019,Saad2020}. In particular, \ac{LSTM} networks are well suited for time-series forecasting~\cite{li2019ea}, making them highly effective for predictive \ac{CSI} in dynamic environments~\cite{Kadambar2023Deep}.

Several deep learning-based predictive \ac{CSI} techniques have been proposed to mitigate the impact of channel aging in fast-fading environments. Some works aim to predict future channel conditions directly from past channel matrices, but this approach often increases model complexity and computational cost. For instance,~\cite{2023Gao} employs Temporal Convolutional Networks (TCNs) to predict future channel states and select the CQI. However, operating directly on full channel matrices results in high dimensionality and complexity. Similarly,~\cite{2021Yuan} introduces a deep learning framework to predict future channel matrices and derive PMI/RI indicators, but it also relies on raw channel coefficients, leading to heavy computational requirements.

A recent approach, Smart-\ac{CSI}~\cite{Kadambar2023Smart}, addresses the prediction problem in \ac{FDD} systems by shifting from the prediction of the channel matrix to the domain of mutual information, thus reducing the dimensionality of the input and the computational load. Nevertheless, Smart-\ac{CSI} operates in \ac{FDD} mode, where downlink \ac{CSI} must be explicitly reported by the \ac{UE}. This requires continuous transmission of reference signals to maintain prediction accuracy, increasing signaling overhead. Moreover, Smart-\ac{CSI} adopts a single fully connected deep neural network architecture and does not explore recurrent or sequential models that may better exploit temporal correlations.

In conventional 5G systems, the mapping between instantaneous SINR values across resource blocks and the selected \ac{MCS} is typically performed using the \ac{EESM} technique~\cite{Lagen2020}. \ac{EESM} compresses the SINR values per subcarrier into a single effective SINR that accurately reflects the overall quality of the link in the bandwidth. This method is both robust and computationally efficient, providing a compact and informative representation of the channel that is well suited for learning-based prediction models.  

Traditional link adaptation commonly employs the outer loop link adaptation (OLLA), a reactive mechanism that adjusts \ac{MCS} based on previous \ac{BLER} outcomes \cite{blanquez2016eolla}. In contrast, predictive approaches proactively estimate future channel conditions, allowing the system to anticipate variations and mitigate performance degradation before transmission errors occur.  

Our work focuses on \ac{TDD} systems, where channel reciprocity eliminates the need for \ac{CSI} feedback, thereby avoiding the signalling overhead inherent to \ac{FDD} solutions such as Smart-\ac{CSI}. Instead of operating on raw channel matrices, we compress the \ac{CSI} into the effective SINR domain using the \ac{EESM} method and evaluate two neural architectures—a fully connected \ac{DNN} and an \ac{LSTM}-based model—to analyse the trade-offs between prediction accuracy, complexity, and latency for real-time deployment in 5G and beyond \ac{TDD} networks.

\subsection*{Contributions}

This work makes the following key contributions:
\begin{itemize}
    \item \textbf{Design of standard-compatible predictors:} We propose both \ac{DNN}- and \ac{LSTM}-based models that operate in the effective \ac{SINR} domain, reducing complexity compared to direct \ac{CSI} matrix prediction and requiring no modifications to the current TDD standard.
    \item \textbf{Comprehensive performance evaluation:} We analyze prediction accuracy (NMSE), throughput, and computational complexity, offering a holistic perspective on predictor performance.
    \item \textbf{Robustness analysis:} We evaluated the generalization of the proposed models across diverse \ac{TDD} propagation conditions, including different TDL channel profiles and both \ac{LOS} and \ac{NLOS} scenarios.
    \item \textbf{System-level validation:} We demonstrate that improvements in prediction accuracy translate into tangible performance gains under channel aging, highlighting the practical benefits of predictive \ac{CSI} for mobility-aware link adaptation.
\end{itemize}

The results provide practical insights into the deployment of AI-driven predictive \ac{CSI} for efficient and reliable link adaptation in \ac{TDD} systems of 5G and beyond.

\section{System Model}

We consider a \ac{MIMO}-\ac{OFDM} system with $N_{\text{Tx}}$ transmit antennas and $N_{\text{Rx}}$ receive antennas, consistent with typical 5G New Radio (NR) configurations operating in \ac{TDD} mode. The subcarrier spacing is set to 15~kHz (numerology 0), and the time–frequency grid is organized into slots of 1~ms duration, each comprising 14 OFDM symbols. The fundamental unit of resource allocation is the \ac{RE}, defined by one subcarrier over one OFDM symbol, while a \ac{RB} consists of 12 contiguous subcarriers. The overall system bandwidth is determined by the number of allocated resource blocks, denoted as $N_{\text{RB}}$.

The wireless channel is modeled using the \ac{TDL} approach, as specified in~\cite{38.901}. This model represents the propagation channel as a superposition of $L$ discrete multipath components, each characterized by a fixed propagation delay and a time-varying complex amplitude. The channel impulse response is expressed as
\begin{equation}
    h(t,\tau) = \sum_{\ell=0}^{L-1} \alpha_{\ell}(t) \, \delta(\tau - \tau_\ell),
\end{equation}
where $\alpha_{\ell}(t)$ denotes the complex gain of the $\ell$-th tap, which implicitly includes the Doppler effects described in~\cite{38.901}. The variable $\tau$ represents the elapsed time since the impulse was applied, corresponding to the delay domain of the channel response, while $\tau_\ell$ is the fixed propagation delay associated with the $\ell$-th multipath component. 

This formulation captures both time-selective and frequency-selective fading, which are intrinsic to wireless propagation environments. In particular, the combination of multipath propagation and user mobility gives rise to \emph{channel aging} in \ac{TDD} systems, where the estimated \ac{CSI} progressively becomes outdated between the estimation instant and its use in downlink transmission.

\subsection{CSI Acquisition Framework}

\begin{figure}[h]
    \centering
    \includegraphics[width=\linewidth]{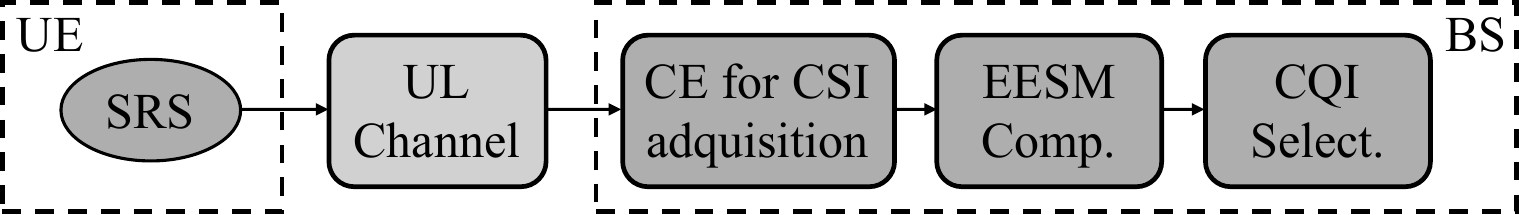}
    \caption{CSI acquisition process in a TDD system.}
    \label{fig:TDDSystem}
\end{figure}

In \ac{TDD} systems, channel reciprocity enables the \ac{BS} to estimate the downlink channel directly from uplink pilot signals, eliminating the need for explicit \ac{CSI} feedback. As illustrated in Fig.~\ref{fig:TDDSystem}, the process begins when the \ac{UE} transmits \ac{SRS}, which propagate through the wireless channel and are received at the \ac{BS}. Exploiting the reciprocal nature of the channel, the \ac{BS} performs downlink channel estimation (CE) based on these uplink measurements.

After the channel has been estimated, the \ac{BS} evaluates the effective channel quality using the \ac{EESM} technique. The resulting effective SINR is then mapped to the most suitable CQI index, which determines the \ac{MCS} used in the next downlink transmission. The \ac{UE} subsequently receives the data, demaps the resource elements, applies channel equalization, and decodes the transmitted information. System performance is commonly assessed in terms of \emph{throughput}, defined as the net rate of successfully delivered information bits.

\begin{figure}[h]
    \centering
    \includegraphics[width=0.95\linewidth]{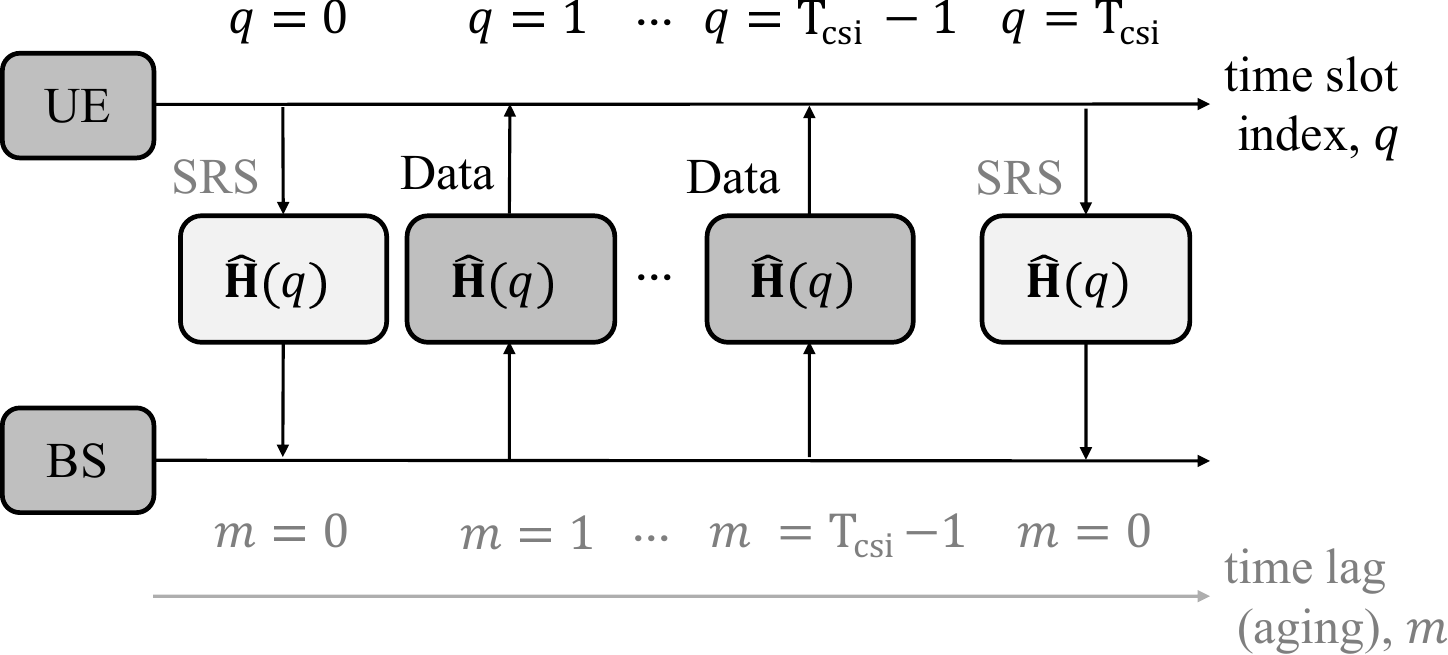}
    \caption{Temporal CSI acquisition and the impact of channel aging.}
    \label{fig:TemporalCSIAdqusition}
\end{figure}

Fig.~\ref{fig:TemporalCSIAdqusition} illustrates the impact of channel aging on \ac{CSI} acquisition in \ac{TDD} systems. At every \( T_{\mathrm{CSI}} \) interval, the \ac{UE} transmits an \ac{SRS}, and the \ac{BS} updates the \ac{MCS} based on the estimated channel quality and the corresponding effective \ac{SINR} obtained through \ac{EESM}. However, as time progresses, this estimate gradually becomes outdated due to user mobility and time-varying fading, thereby requiring new reference signals to sustain accurate link adaptation. 

The estimated channel used for data transmission at slot \( q \) is given by
\begin{equation}
    \hat{\mathbf{H}}(q) = \mathbf{H}\!\left( \left\lfloor \frac{q}{T_{\mathrm{CSI}}} \right\rfloor \!\cdot\! T_{\mathrm{CSI}} \right),
\end{equation}
where \( \mathbf{H} \) denotes the true channel matrix. The variable \( m \) in Fig.~\ref{fig:TemporalCSIAdqusition} represents the elapsed time, measured in slots, since the last channel estimate was obtained. As \( m \) increases within the reporting interval, the estimated channel \(\hat{\mathbf{H}}(q)\) progressively diverges from the true channel \(\mathbf{H}(q)\), demonstrating the obsolescence of the reported \ac{CSI} and the resulting degradation in link adaptation accuracy.

During \ac{CSI} acquisition, the \ac{SINR} is computed for each \ac{RB} and spatial layer. To consolidate this multidimensional information into a single representative metric for link adaptation, the \ac{EESM} method is applied jointly across all layers and RBs. For each candidate \ac{CQI} level $i$, the effective SINR is calculated as
\begin{equation}
    \gamma^{(i)}_{\mathrm{eff}} = -\beta^{(i)} \ln \left( \frac{1}{N_{\mathrm{L}} N_{\mathrm{RB}}} \sum_{l=1}^{N_{\mathrm{L}}} \sum_{n=1}^{N_{\mathrm{RB}}} \exp\left( -\frac{\gamma(l,n)}{\beta^{(i)}} \right) \right),
\end{equation}
where $\gamma(l,n)$ is the instantaneous SINR at the $l$-th spatial layer and $n$-th resource block, $N_{\mathrm{L}}$ is the number of layers, and $\beta^{(i)}$ is a \ac{CQI}-dependent calibration parameter.

The optimal \ac{CQI} index is then selected as
\begin{equation}
    i^* = \max \left\{ i \,:\, \mathrm{BLER}\!\left( \gamma^{(i)}_{\mathrm{eff}} \right) \leq \mathrm{BLER}_{\mathrm{target}} \right\},
\end{equation}
thus maximizing spectral efficiency while satisfying the target reliability constraint.  

For the prediction framework, we use the effective SINR $\gamma^{(i^*)}_{\mathrm{eff}}$ associated with the selected \ac{CQI} value. For readability, this value is denoted simply as $\gamma_{\mathrm{eff}}$ in the following sections. Each selected \ac{CQI} corresponds to a specific \ac{MCS}, which defines the spectral efficiency according to the 3GPP specification~\cite[Table~5.2.2.1-2]{38.214}. Consequently, the effective SINR predicted by the proposed model directly determines the achievable throughput, linking the quality of the \ac{CSI} prediction to the overall system performance.

\section{Proposed \ac{CSI} Prediction Framework}

The temporal evolution of wireless channels generally exhibits correlation over time, enabling the inference of future channel states from past observations. In \ac{TDD} systems, where channel reciprocity allows the \ac{BS} to obtain accurate downlink \ac{CSI} directly from uplink pilots, this property can be exploited to forecast channel quality beyond the current reporting interval.

\begin{figure}[h]
    \centering
    \includegraphics[width=0.8\linewidth]{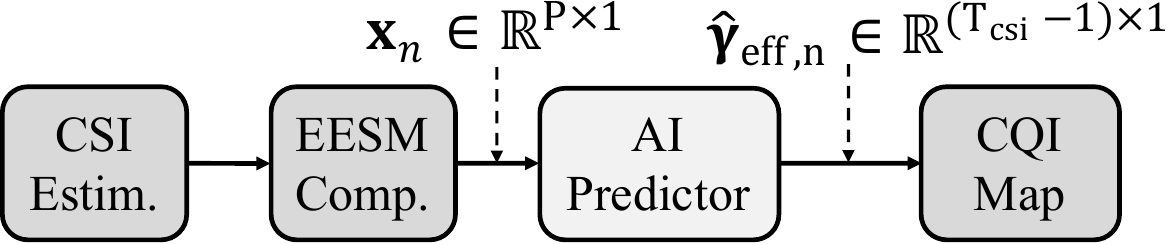}
    \caption{CSI prediction framework.}
    \label{fig:PredictionFramework}
\end{figure}

Figure~\ref{fig:PredictionFramework} illustrates the overall \ac{CSI} prediction framework. The process begins with \ac{CSI} estimation and compression through \ac{EESM}, which generates the input vector \( \mathbf{x}_n \). At each reporting instant \( n \), the input to the predictor is defined as
\begin{equation}
    \mathbf{x}_n = 
       \left[ \gamma_{\mathrm{eff}}(n), \gamma_{\mathrm{eff}}(n - T_{\mathrm{CSI}}), \ldots, \gamma_{\mathrm{eff}}(n - P \cdot T_{\mathrm{CSI}}) \right]^\top,
\end{equation}
where \( P \) denotes the prediction window length and \( T_{\mathrm{CSI}} \) is the \ac{CSI} reporting interval, measured in slots. This vector captures the temporal dynamics of the channel quality and serves as the input sequence to the proposed learning-based models.

This vector is then processed by one of the two considered neural architectures: a fully connected \ac{DNN} or a recurrent network with \ac{LSTM} units. The predicted effective SINR values define the vector of predicted SINRs as
\begin{equation}
\hat{\boldsymbol{\gamma}}_{\mathrm{eff}, n} = 
\left[ \hat{\gamma}_{\mathrm{eff}}(n+1), \hat{\gamma}_{\mathrm{eff}}(n+2), \ldots, \hat{\gamma}_{\mathrm{eff}}(n+T_{\mathrm{CSI}}-1)
\right]^\top,
\label{eq:outputSINR}
\end{equation}
which are finally mapped to the corresponding CQI levels, enabling proactive link adaptation.

Importantly, the proposed framework is fully compatible with current 5G standards. Since it only relies on information obtained from periodic \ac{CSI} reporting based on reference signals, no additional pilots or signalling overhead are required. Unlike alternative approaches that assume continuous \ac{CSI} measurements, our method operates with the same feedback already available in practical systems, ensuring seamless integration with existing deployments.

\subsection{DNN-Based Prediction Network}

As a baseline, we implement a fully connected \ac{DNN} with a single hidden layer that learns a nonlinear mapping between past and future SINR values. At the reporting instant \( n \), the input to the network is the vector \( \mathbf{x}_n \), defined in (1).

The hidden layer applies an affine transformation followed by a nonlinear activation function:
\begin{equation}
    \mathbf{h} = \sigma \left( \mathbf{W}^{(1)} \mathbf{x}_n + \mathbf{b}^{(1)} \right),
\end{equation}
where \( \mathbf{W}^{(1)} \) and \( \mathbf{b}^{(1)} \) denote the weight matrix and bias vector, respectively, and \( \sigma(\cdot) \) is a nonlinear activation function, such as ReLU.

The output layer produces the predicted effective SINR values for the next $T_{\mathrm{CSI}}$ slots:
\begin{equation}
\hat{\boldsymbol{\gamma}}_{\mathrm{eff},n} = 
\mathbf{W}^{(2)} \mathbf{h} + \mathbf{b}^{(2)},
\end{equation}
where \(\mathbf{h}\) is the hidden layer representation, and \(\mathbf{W}^{(2)}\) and \(\mathbf{b}^{(2)}\) are the weight matrix and bias vector of the output layer, respectively. The model is trained to minimize the mean squared error (MSE), $\mathcal{L}$ as
\begin{equation}
    \mathcal{L} = \frac{1}{T_{\mathrm{CSI}}} \sum_{k=1}^{T_{\mathrm{CSI}}} 
    \left( \gamma_{\mathrm{eff}}(n+k) - \hat{\gamma}_{\mathrm{eff}}(n+k) \right)^2.
\end{equation}

\begin{figure}[h]
    \centering
    \subfloat[]{
        \includegraphics[width=0.45\linewidth]{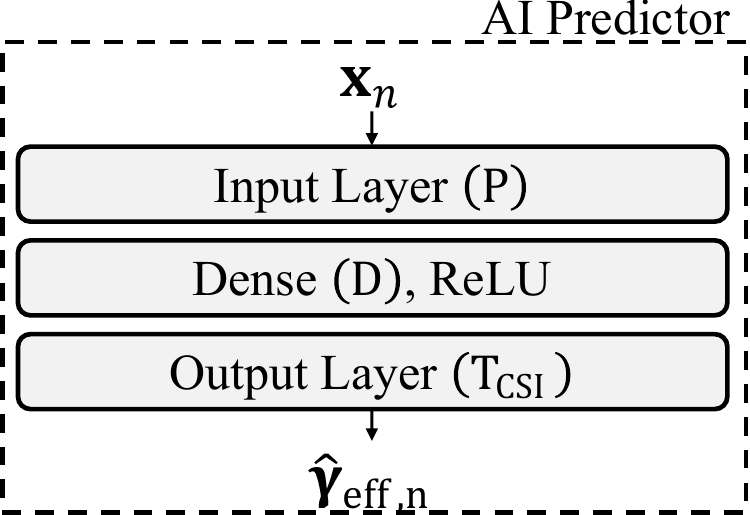}
        \label{fig:PredictionFrameworkDNN}
    }
    \hfill
    \subfloat[]{
        \includegraphics[width=0.45\linewidth]{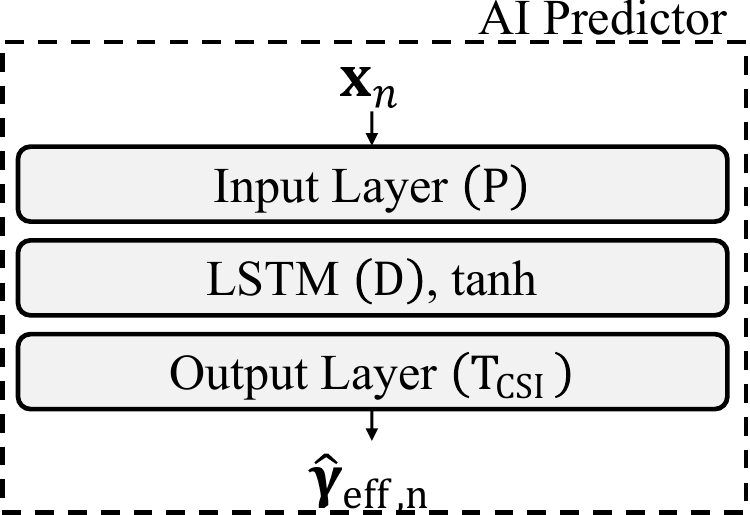}
        \label{fig:PredictionFrameworkLSTM}
    }
    \caption{Proposed CSI prediction architectures corresponding to the AI Predictor block in Fig.~\ref{fig:PredictionFramework}: (a) fully connected DNN and (b) LSTM-based network.}
    \label{fig:PredictionArchitectures}
\end{figure}

Figure~\ref{fig:PredictionFrameworkDNN} illustrates the \ac{DNN} predictor. The architecture consists of an input layer of dimension \( \mathrm{P} \), corresponding to the input vector \( \mathbf{x}_n \), followed by a single hidden layer of dimension \( \mathrm{D} \) with ReLU activation, and an output layer of dimension \( T_{\mathrm{CSI}} \), representing the predicted effective SINR values.

\subsection{LSTM-Based Prediction Network}

To capture temporal correlations more effectively, we employ a recurrent neural network with \ac{LSTM} units. Unlike the \ac{DNN}, which processes past samples in a static manner, the \ac{LSTM} maintains internal states that allow it to exploit both short- and long-term dependencies in the SINR sequence.

\begin{figure}[h]
    \centering
    \includegraphics[width=0.85\linewidth]{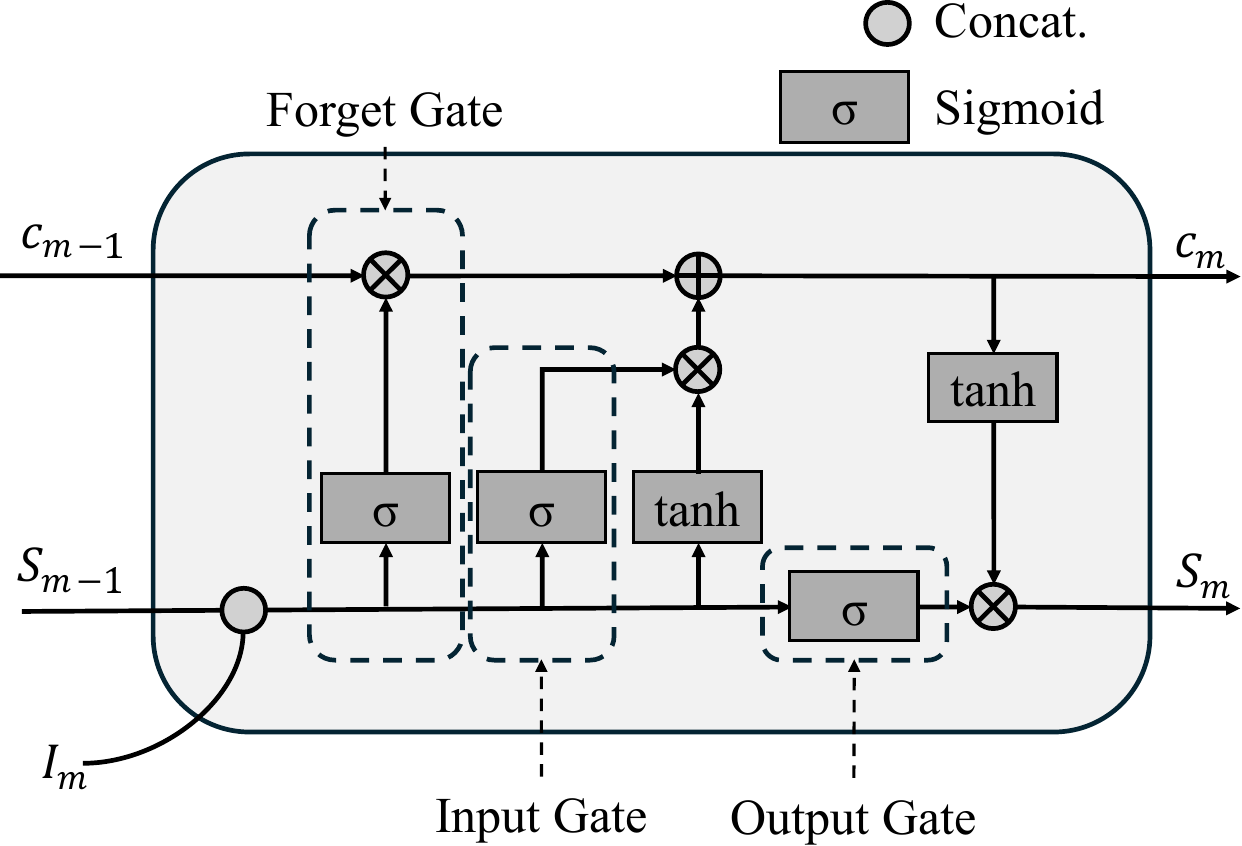}
    \caption{Structure of an LSTM unit.}
    \label{fig:LSTMCell}
\end{figure}

An \ac{LSTM} cell incorporates three gating mechanisms: the input, forget, and output gates, each controlled by a sigmoid activation function \cite{Mattu2022}. These gates regulate the flow of information across time steps, enabling the network to retain relevant temporal patterns while discarding obsolete ones. The structure of an individual \ac{LSTM} unit is shown in Fig.~\ref{fig:LSTMCell}.

The overall structure of the \ac{LSTM}-based predictor is illustrated in Fig.~\ref{fig:PredictionFrameworkLSTM}. The input layer is identical to that of the \ac{DNN}-based model. However, the hidden layer is implemented as an \ac{LSTM} layer of dimension $\mathrm{D}$. This parameter is varied in the experiments to evaluate its impact on complexity, number of hidden states, and predictive performance. The \ac{LSTM} employs $\tanh$ activation functions. Finally, the output layer has $T_{\mathrm{CSI}}$ dimensions, corresponding to the predicted effective SINR values.

This design enables the \ac{BS} in \ac{TDD} systems to proactively select the most spectrally efficient \ac{MCS} for upcoming transmissions, thereby improving throughput and reliability while avoiding explicit \ac{CSI} feedback from the \ac{UE}.

\section{Simulation Results}

This section evaluates the performance of the proposed \ac{CSI} prediction frameworks through a series of simulation-based experiments. Two complementary performance metrics are analyzed:
\begin{itemize}
    \item Normalized Mean Square Error (NMSE): quantifies the accuracy of the predicted effective SINR in \ref{eq:outputSINR}.
    \item Throughput: measures the system-level impact of prediction on link adaptation and data transmission efficiency.
\end{itemize}

The comparison is conducted between the \ac{LSTM}-based predictor and a fully connected \ac{DNN} baseline. The simulation parameters are summarized in Table~\ref{tab:SimulationParameters}.

\begin{table}[h]
    \centering
    \begin{tabular}{c|c}
        \hline
        \textbf{Parameter}  & \textbf{Value} \\
        \hline
        Average SNR & 12.5 dB \\
        \hline
        Subcarrier Spacing (SCS) & 15 kHz \\
        \hline
        Number of RBs ($N_{\mathrm{RB}}$) & 52 \\
        \hline
        Bandwidth & 10 MHz \\
        \hline
        Maximum Doppler Shift ($f_{\textrm{D}}$) & [1--30] Hz \\
        \hline
        Delay Spread & 300 ns \\
        \hline
        MIMO Configuration & $4 \times 4$ \\
        \hline
        Transmission Layers & 4 \\
        \hline
        Channel Models & TDL-A / TDL-D \\
        \hline
        CSI Reporting Period ($T_{\textrm{CSI}}$) & 4 slots \\
        \hline
        Transmission Interval & 1 ms \\
        \hline
    \end{tabular}
    \caption{Simulation parameters.}
    \label{tab:SimulationParameters}
\end{table}

A dataset of 510{,}000 samples was generated, split into 400{,}000 for training, 100{,}000 for validation, and 10{,}000 for testing. Both models were trained for 200 epochs with a batch size of 256, ensuring convergence under diverse channel conditions.

\subsection{Prediction Accuracy: NMSE Evaluation}

\begin{figure}[h]
    \centering
    \includegraphics[width=0.95\linewidth]{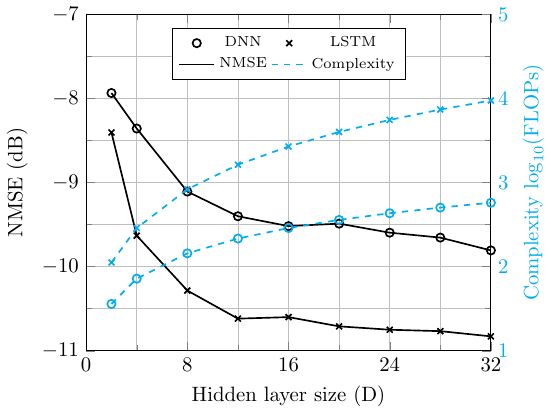}
    \caption{Trade-off between NMSE (dB) and computational complexity (FLOPs) for different hidden layer sizes in \ac{LSTM} and DNN predictors at Doppler = 10 Hz in a TDL-A channel for $T_{\mathrm{CSI}} = 4$}.
    \label{fig:NMSEvsComplexity}
\end{figure}

We first evaluate the prediction accuracy of the models in terms of NMSE. Fig.~\ref{fig:NMSEvsComplexity} illustrates the trade-off between NMSE and computational complexity for hidden layer sizes ($\mathrm{D}$) ranging from 2 to 32 units. The right y-axis (cyan) represents the computational complexity in floating-point operations (FLOPs) in logarithmic scale, while the left y-axis (black) shows the NMSE in dB. As expected, increasing $\mathrm{D}$ improves prediction accuracy at the cost of higher computational demand. For instance, at $\mathrm{D}=16$, the NMSE decreases by approximately 1.1~dB, whereas the complexity increases by about 2400 FLOPs.

Based on this analysis, $\mathrm{D}=16$ was selected as the default hidden size for both predictors, as it provides a good compromise between accuracy and complexity, ensuring a fair comparison.

\begin{figure}[h]
    \centering
    \includegraphics[width=0.95\linewidth]{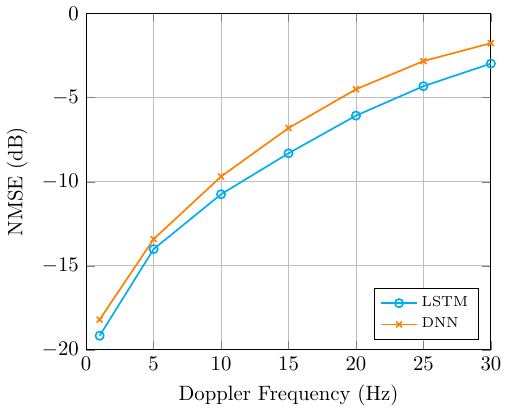}
    \caption{NMSE performance of LSTM and DNN predictors versus Doppler frequency for \(T_{\mathrm{CSI}} = 4\) in a TDL-A channel.}
    \label{fig:NMSESweepDopplerTDL-A}
\end{figure}

Fig.~\ref{fig:NMSESweepDopplerTDL-A} shows the performance of NMSE as a function of the Doppler frequency for $T_{\mathrm{CSI}} = 4$ on a TDL-A channel. At low Doppler values (1--5~Hz), both models achieve similar accuracy, with less than 1~dB difference. At moderate Doppler values (10--20~Hz), the \ac{LSTM} clearly outperforms the \ac{DNN}, reducing the NMSE by up to 2~dB. At higher Doppler values, the NMSE of both models converges toward 0~dB, reflecting the difficulty of reliable prediction under very fast channel dynamics.

\begin{figure}[h]
    \centering
    \includegraphics[width=0.95\linewidth]{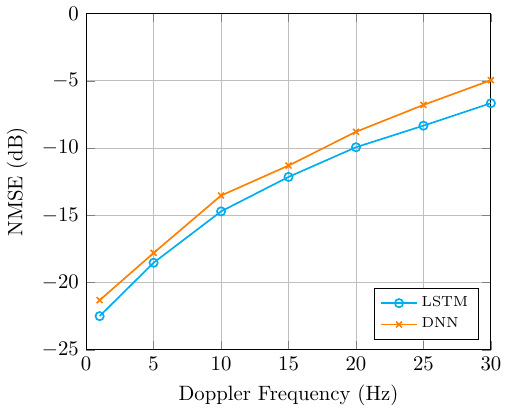}
    \caption{NMSE performance of LSTM and DNN predictors versus Doppler frequency for \(T_{\mathrm{CSI}} = 4\) in a TDL-D channel.}
    \label{fig:NMSESweepDopplerTDL-D}
\end{figure}

Fig.~\ref{fig:NMSESweepDopplerTDL-D} extends this analysis to a TDL-D channel, characterized by the presence of a \ac{LOS} component. Compared to TDL-A, both techniques achieve improved robustness, as the deterministic \ac{LOS} path reduces channel variability. Across the entire Doppler range, the \ac{LSTM} consistently outperforms the \ac{DNN} by about 1~dB, confirming its superior ability to exploit temporal correlations in mixed \ac{LOS} and multipath conditions. These results demonstrate that both predictors generalize effectively across different propagation environments.

\subsection{System-Level Impact: Throughput Evaluation}

While NMSE quantifies the accuracy of the predicted \ac{CSI}, the ultimate goal is to enhance system-level performance. To this end, we evaluate the throughput, defined as the average successfully delivered data rate, accounting for retransmissions and error correction. This metric captures the practical benefits of predictive \ac{CSI} in enabling more accurate and efficient link adaptation.

The evaluation is conducted using a link-level simulator implemented in MATLAB, which models the complete transmission chain—including channel estimation, CQI selection, modulation, coding, and decoding. The throughput is computed as the average rate of correctly decoded transport blocks, excluding erroneous transmissions from the total transmitted data. This approach provides a realistic assessment of how prediction accuracy translates into effective data rate improvements under varying Doppler and channel conditions.

Prediction accuracy critically determines the balance between reliability and spectral efficiency. An overly optimistic predictor may select an excessively high \ac{MCS}, increasing error rates and retransmissions, whereas a conservative predictor selects lower \ac{MCS} levels, sacrificing throughput. Therefore, accurate \ac{CSI} prediction is essential to maintain an optimal trade-off between throughput and reliability in dynamic wireless environments.

\begin{figure}[h]
    \centering
    \includegraphics[width=0.95\linewidth]{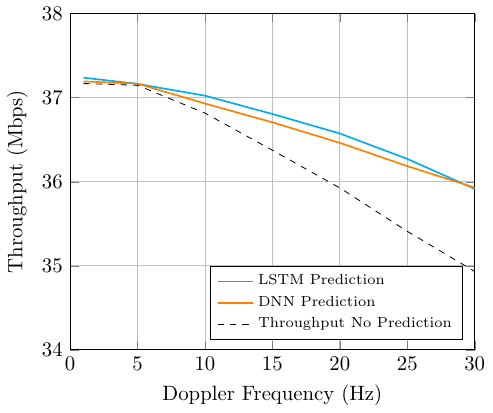}
    \caption{Throughput performance of LSTM and DNN predictors versus Doppler frequency compared to a no-prediction baseline, with \( T_{\mathrm{CSI}} = 4 \) slots.}
    \label{fig:ThroughputSweepDoppler}
\end{figure}

Fig.~\ref{fig:ThroughputSweepDoppler} summarizes the system-level performance as a function of Doppler frequency. At low Doppler values (below 5~Hz), both predictors yield modest improvements of approximately 0.2--0.3~Mbps compared to the no-prediction baseline, since \ac{CSI} variations are minimal. As Doppler increases to moderate levels (10--20~Hz), prediction becomes more relevant: the \ac{DNN} achieves throughput gains of 0.5--0.7~Mbps, while the LSTM attains slightly higher gains, around 0.8--1.0~Mbps. At high Doppler frequencies (25--30~Hz), both predictors maintain improvements, with the LSTM reaching up to 1.2~Mbps, confirming the robustness of the proposed predictive approach under fast time-varying channels.

Overall, the results demonstrate that predictive \ac{CSI} effectively mitigates channel aging and enhances link adaptation in \ac{TDD} systems. Both neural architectures deliver consistent throughput improvements without introducing additional signalling overhead. While the LSTM shows slightly higher resilience at higher Doppler rates, both models achieve favorable trade-offs between accuracy, complexity, and reliability, making them practical solutions for real-time link adaptation in mobility-aware 5G and beyond networks.

\section{Conclusions}

This work has addressed the challenge of channel aging in \ac{TDD} systems by proposing a predictive framework that estimates future \ac{CSI} values in the effective SINR domain. Unlike traditional approaches that either rely on frequent pilot transmissions or operate directly on high-dimensional \ac{CSI} matrices, the proposed solution leverages standard-compliant reference signal reports. This design avoids additional signalling overhead and ensures compatibility with existing 5G systems.

Two neural predictors were developed and evaluated: a fully connected \ac{DNN} and an LSTM network. Simulation results confirmed that both models can exploit temporal channel correlations to improve \ac{CSI} prediction. Among them, the LSTM consistently achieved superior performance, reducing NMSE and delivering higher throughput across a wide range of Doppler frequencies. Notably, the LSTM predictor provided throughput gains of up to 1.2~Mbps compared to a no-prediction baseline, clearly demonstrating its effectiveness in mitigating the adverse effects of channel aging and enhancing link adaptation.

The results highlight that predictive \ac{CSI}, even when implemented with compact neural architectures, can significantly improve spectral efficiency and reliability in mobility scenarios. This makes it a promising candidate for real-time deployment in 5G and beyond networks, where maintaining robust performance under user mobility is critical.

\section*{Acknowledgment}
This work has been funded by MCIN/AEI/10.13039/501100011033 (Spain), by the European Fund for Regional Development (FEDER) {`\textit{A way of making Europe}´}, Keysight Technologies, and the University of M\'alaga through grants PID2020-118139RB-I00, RYC2021-034620-I and 8.06/6.10.6635.


\bibliographystyle{IEEEtran}
\bibliography{references.bib}

\begin{thebibliography}{10}
\providecommand{\url}[1]{#1}
\csname url@samestyle\endcsname
\providecommand{\newblock}{\relax}
\providecommand{\bibinfo}[2]{#2}
\providecommand{\BIBentrySTDinterwordspacing}{\spaceskip=0pt\relax}
\providecommand{\BIBentryALTinterwordstretchfactor}{4}
\providecommand{\BIBentryALTinterwordspacing}{\spaceskip=\fontdimen2\font plus
\BIBentryALTinterwordstretchfactor\fontdimen3\font minus \fontdimen4\font\relax}
\providecommand{\BIBforeignlanguage}[2]{{%
\expandafter\ifx\csname l@#1\endcsname\relax
\typeout{** WARNING: IEEEtran.bst: No hyphenation pattern has been}%
\typeout{** loaded for the language `#1'. Using the pattern for}%
\typeout{** the default language instead.}%
\else
\language=\csname l@#1\endcsname
\fi
#2}}
\providecommand{\BIBdecl}{\relax}
\BIBdecl

\bibitem{Vega2021}
F.~J. Martín-Vega, J.~C. Ruiz-Sicilia, M.~C. Aguayo, and G.~Gómez, ``Emerging tools for link adaptation on {5G NR} and beyond: Challenges and opportunities,'' \emph{IEEE Access}, vol.~9, pp. 126\,976--126\,987, 2021.

\bibitem{Goldsmith2005}
A.~Goldsmith, \emph{Wireless Communications}.\hskip 1em plus 0.5em minus 0.4em\relax Cambridge university press, 2005.

\bibitem{Papazafeiropoulos2017}
A.~K. Papazafeiropoulos, ``Impact of general channel aging conditions on the downlink performance of {Massive MIMO},'' \emph{IEEE Transactions on Vehicular Technology}, vol.~66, no.~2, pp. 1428--1442, 2017.

\bibitem{Truong2013}
K.~T. Truong and R.~W. Heath, ``Effects of channel aging in massive {MIMO} systems,'' \emph{Journal of Communications and Networks}, vol.~15, no.~4, pp. 338--351, 2013.

\bibitem{Jiang2017}
C.~Jiang, H.~Zhang, Y.~Ren, Z.~Han, K.-C. Chen, and L.~Hanzo, ``Machine learning paradigms for next-generation wireless networks,'' \emph{IEEE Wireless Communications}, vol.~24, no.~2, pp. 98--105, 2017.

\bibitem{Ye2018}
H.~Ye, G.~Y. Li, and B.-H. Juang, ``Power of deep learning for channel estimation and signal detection in {OFDM} systems,'' \emph{IEEE Wireless Communications Letters}, vol.~7, no.~1, pp. 114--117, 2018.

\bibitem{Liao2019}
Y.~Liao and et~al, ``{CSI} feedback based on deep learning for {Massive MIMO} systems,'' \emph{IEEE Access}, vol.~7, pp. 86\,810--86\,820, 2019.

\bibitem{Saad2020}
W.~Saad, M.~Bennis, and M.~Chen, ``A vision of {6G} wireless systems: Applications, trends, technologies, and open research problems,'' \emph{IEEE Network}, vol.~34, no.~3, pp. 134--142, 2020.

\bibitem{li2019ea}
Y.~Li, Z.~Zhu, D.~Kong, H.~Han, and Y.~Zhao, ``{EA-LSTM}: Evolutionary attention-based {LSTM} for time series prediction,'' \emph{Knowledge-Based Systems}, vol. 181, p. 104785, 2019.

\bibitem{Kadambar2023Deep}
S.~Kadambar and et~al, ``Deep learning based joint {CSI} compression and prediction for beyond-{5G} systems,'' in \emph{Proc. IEEE Global Communications Conference}, 2023, pp. 4792--4797.

\bibitem{2023Gao}
J.~Gao and et~al, ``Fast time-varying wireless channel prediction based on deep learning,'' in \emph{Proc. 9th International Conference on Computer and Communications (ICCC)}, 2023, pp. 940--945.

\bibitem{2021Yuan}
Z.~Yuan, K.~Niu, and C.~Dong, ``Channel prediction and {PMI/RI} selection in {MIMO-OFDM} systems based on deep learning,'' in \emph{Proc. IEEE 32nd Annual International Symposium on Personal, Indoor and Mobile Radio Communications (PIMRC)}, 2021, pp. 598--603.

\bibitem{Kadambar2023Smart}
S.~Kadambar and et~al, ``Smart-{CSI}: Deep learning based low complexity {CSI} prediction for beyond-{5G} systems,'' in \emph{Proc. IEEE 98th Vehicular Technology Conference (VTC2023-Fall)}, 2023, pp. 1--5.

\bibitem{Lagen2020}
S.~Lagen and et~al, ``New radio physical layer abstraction for system-level simulations of {5G} networks,'' in \emph{Proc. IEEE International Conference on Communications (ICC)}, 2020, pp. 1--7.

\bibitem{blanquez2016eolla}
F.~Blanquez-Casado, G.~Gomez, M.~d.~C. Aguayo-Torres, and J.~T. Entrambasaguas, ``{eOLLA: an enhanced outer loop link adaptation for cellular networks},'' \emph{EURASIP Journal on Wireless Communications and Networking}, vol. 2016, no.~1, p.~20, 2016.

\bibitem{38.901}
3GPP, \emph{{Technical Report (TR); Study on channel model for frequencies from 0.5 to 100 GHz}}, {3rd Generation Partnership Project (3GPP)} TR {38.901}, Rev. 17.0.0, April 2022.

\bibitem{38.214}
------, \emph{{Technical Specification (TS); Physical layer procedures for data}}, {3rd Generation Partnership Project (3GPP)} TS {38.214}, Rev. 17.12.0, January 2025.

\bibitem{Mattu2022}
S.~R. Mattu and et~al, ``{Deep Channel Prediction: A DNN Framework for Receiver Design in Time-Varying Fading Channels},'' \emph{IEEE Transactions on Vehicular Technology}, vol.~71, no.~6, pp. 6439--6453, 2022.

\end{thebibliography}
\end{document}